\documentclass[aps,graphics,twocolumn,showpacs]{revtex4}
\usepackage{graphicx}
\usepackage{verbatim}
\usepackage{amssymb}
\usepackage{color}
\providecommand{\boldsymbol}[1]{\mbox{\boldmath $#1$}}
\newcommand{\y}{y}

\begin{document}
\title{Locality, detection efficiencies, and probability polytopes}
\author{J. Wilms$^{1,2}$, Y. Disser$^{2}$, G. Alber$^{2}$, I. C. Percival$^{3, 4}$}
\affiliation{$^{1}$Fakult\"at f\"ur Physik, Universit\"at Wien, 1090 Wien, Austria\\
$^{2}$Institut f\"ur Angewandte Physik, Technische Universit\"at Darmstadt, 64289 Darmstadt, Germany\\
$^{3}$Department of Physics, Queen Mary, University of London, London E1 4NS, United Kingdom\\
$^{4}$Department of Physics and Astronomy, University of Sussex, Brighton BN1 9QH, United Kingdom}
\date{\today}

\begin{abstract}
We present a detailed investigation of minimum detection efficiencies,
below which locality cannot be violated by any quantum system of any
dimension in bipartite Bell experiments.  Lower bounds on these
minimum detection efficiencies are determined with the help of linear
programming techniques.  Our approach is based on the observation that
any possible bipartite quantum correlation originating from a quantum
state in an arbitrary dimensional Hilbert space is sandwiched between
two probability polytopes, namely the local (Bell) polytope and a
corresponding nonlocal no-signaling polytope.  Numerical results are
presented demonstrating the dependence of these lower bounds on the
numbers of inputs and outputs of the bipartite physical system.
\end{abstract}
\maketitle

\section{Introduction}

Despite numerous recent experimental tests for violations of locality
by quantum theory, such as the experiment by Weihs et
al. \cite{Bellexperiments}, we still do not know for certain whether
or not the laws of physics are entirely local \cite{gisinbell}.
This is because so far
no single experiment has closed both the locality and the detection
loophole simultaneously.  

In general, classical correlations between two spacelike separated
experimenters, Alice (A) and Bob (B), obey locality constraints, which
can be expressed in terms of (generalized) Bell inequalities.
According to Bell's theorem \cite{Speakable}, these inequalities can
be violated if the relevant correlations are produced by ideal
measurements of an entangled quantum system, whose quantum state is
required to originate in the common backward light cone of A and B.
This is weak nonlocality.  Strong nonlocality would be the
non-existent correlations produced by signaling faster than the speed
of light.

In a typical two-photon Bell experiment the polarization state of a
pair of entangled photons is measured independently by A and B.
Preferred states for such experiments are pure two-photon states of maximum
entanglement, the so-called Bell states.

One of the last remaining major problems on the way to a loophole-free test of
locality is the detection loophole, which comes from photon detection
efficiencies being too small \cite{Cabello,Gisin}.  In this context
Eberhard \cite{Eberhard} recognized that weakly entangled pure
two-photon states yield a maximized tolerance to detection
inefficiency.  Using numerical optimization he was thus able to reduce
the critical detection efficiency in the two-photon Bell experiment
towards a theoretical limit of 2/3 under the assumption of identical
detection efficiencies for A and B.  Later authors
\cite{Cabello,Gisin} have applied his method to other experiments of
the Bell type and achieved theoretical values of the critical
detection efficiency as low as $0.43$ for the extreme asymmetric case
in which only either A's or B's detection is perfect.

In view of these results on minimum detection efficiencies
two major questions arise.  Firstly, it is not clear whether these
results also apply to Bell experiments in which the dichotomic
variables measured by A and B result from quantum observables and
quantum states in arbitrarily high dimensional Hilbert
spaces. Secondly, it is unclear what influence symmetric and
asymmetric detection efficiencies have on cases in which more than two
physical quantities are measured on A's and B's sides or on cases in
which the observables have more than two possible outcomes. It is the
main purpose of this paper to address these open questions.

For this purpose an efficient way is developed for describing local bipartite
correlations with the help of probability polytopes and 
linear programming. It is known that the relevant
local polytopes can be described efficiently in terms of their
vertexes, which can be obtained for any experimental setup of any
number of inputs and outputs, as described
here in Sec.~\ref{CCPP}. Thus, any test of locality reduces to an
inclusion test determining whether a given set of probabilities is
located outside or inside the relevant local polytope.  In addition,
with the help of a second class of probability polytopes which
describe nonlocal no-signaling correlations \cite{resource} it is
possible to obtain lower bounds on minimum detection efficiencies for
bipartite Bell experiments. These latter probability polytopes include
all correlations of bipartite quantum systems of any dimension and
thus yield dimension-independent lower bounds on detection
efficiencies.  First results of such lower bounds are presented for
inefficiencies of arbitrary symmetry and for bipartite locality tests
with dichotomic variables which involve random choices of A's and B's
observables from a set of up to four elements.

In addition, some new results for lower bounds on higher numbers of
outputs are presented.  Finally, it is demonstrated that the $1$-norm
(used here as the distance) between the point defined by the observed
probabilities and the relevant local polytope represents a
convenient way of quantifying violations of locality in the presence
of experimental uncertainties.  This distance can be determined in a
straightforward way by linear programming.

This paper is organized as follows: Sec.~\ref{CCPP} summarizes
relevant and already known results on classical correlations,
classical transfer functions, and their relation to probability
polytopes.  The local polytopes and nonlocal no-signaling polytopes
are introduced. These describe classical local correlations, and
classical nonlocal correlations which fulfill the no-signaling
condition, respectively. It is shown how these polytopes can be
represented in terms of their vertexes or equivalently in terms of
inequalities for their facets.

In Sec.~\ref{Detineff} these two types of polytope are used to
determine lower bounds on minimum detection efficiencies which still
allow for a violation of locality by quantum systems.
Sec.~\ref{Distmeas} finally demonstrates how the $1$-norm defining the
distance of a given probability distribution from the relevant local
polytope can be determined with the help of linear programming.

\section{Classical correlations and probability polytopes\label{CCPP}}

In this section basic concepts involved in the description of
classical bipartite correlations are summarized.  For this purpose
transfer functions and probability polytopes are introduced
\cite{percival1998qtf,percival1999qmb,Ziegler(1995)}.  In particular, the local Bell polytope
${\cal L}$ and the nonlocal no-signaling polytope ${\cal P}$ are discussed in
detail.

\subsection{Transfer functions and transition probabilities\label{sec:polytopes}}
Given a classical deterministic system with discrete inputs $x$ and
outputs $a$, the output is a definite function $F$ of the input:
$a=F(x)$. Thereby the transfer function of the system, $F$, specifies a
single transition from $x$ to $a$ for every input $x$.  If the system
may be stochastic, then the behavior of the system has to be described
in terms of the transition probabilities $P(a|x)$, which define a
point in a transition probability space whose coordinates are these
probabilities.  Since, for a given input, the total probability of an
output must be unity, the probabilities satisfy the normalization
condition $\sum_a P(a|x)=1$.  
For a deterministic system with transfer
function $F$ the probabilities are $P(a|x) = \delta(a,F(x))$, with
possible values 0 or 1. In terms of these particular probabilities
an arbitrary transition probability of a stochastic system can be represented
by \cite{percival1999qmb}
\begin{eqnarray}\label{Pax}
P(a|x)&=&\sum_F P(F)\delta(a,F(x))
\end{eqnarray}
with $P(F)$ denoting the probability with which the deterministic transfer function $F$
governs the correlations under consideration.
 
If there are $N(x)$ possible values for the input $x$ and $N(a)$ possible
values for the output $a$, there are $N(a)^{N(x)}$ possible transfer
functions, but only $N(x)\times N(a)$ transition probabilities, so usually there are
many more transfer functions than there are transition
probabilities and the expansion in terms of transfer functions is not
generally unique.  The sum of Eq.(\ref{Pax}) is over all transfer functions, but if
there are constraints on them, it can be over a subset of $F$. 

A typical bipartite Bell experiment testing locality, such as the
one described by Eberhard \cite{Eberhard}, involves a two-photon
source distributing one photon to Alice (A) and the other photon to
Bob (B).  The subsequent experiment performed by A and B may be
considered as an input-output system, in which A's input $x$ is a
choice of angle for the measurement of a photon polarization and her
output $a$ is the result of the measurement, $+$ or $-$, depending on
whether the polarization is found to be parallel or perpendicular to
the chosen angle.  Similarly for B with input $y$ and output $b$.
In the simplest case A and B each have a choice of two angles
only, a different pair for A and for B.  So each of them has 2
inputs and 2 outputs, resulting in 4 inputs and 4 outputs for the whole
system.  In generalizations of bipartite Bell experiments the number
of inputs as well as the number of possible outputs of A and B
may also be larger.  Notice that the outputs are classical events
which result from quantum measurements.  Since the transition
probabilities are all probabilities of these classical events, the
analysis of a Bell experiment does not depend in any way on quantum
theory, although the design of such an experiment clearly does.

Assuming locality means that for deterministic systems A's output can
only depend on her input, and the same for B.  So a transfer function
$F$ for the whole system is made up of one transfer function for A and
one for B: $F=(F^A,F^B)$, where $a=F^A(x)$ and $b=F^B(y)$.  This is
the locality constraint on transfer functions, which in general
reduces their possible number significantly.  

Thus if the numbers of possible inputs and outputs of A are denoted
by $N(x)$ and $N(a)$ and of B by $N(y)$ and $N(b)$ respectively, the
total number of local transfer functions is given by
$N(a)^{N(x)}\times N(b)^{N(y)}$ and the number of corresponding local
transition probabilities is given by $N(x)\times N(a)\times N(y)\times
N(b)$, although the latter are not independent.  

So for any classical theory of a Bell experiment, the transition
probabilities must be obtainable from some local transfer function
probabilities using a basic equation of the form
\begin{eqnarray}
P(ab|xy)&=&\sum_{F^A,F^B} P(F^A,F^B)\delta(a,F^A(x))\delta(b,F^B(y)).\nonumber\\
&&
\label{basicequation}
\end{eqnarray}

\subsection{Probability polytopes and their representations}

A convex polytope in a space of dimension $D$ is a generalization of a
convex polygon in 2-space or of a convex polyhedron in 3-space.  It
can be defined as all those points whose coordinates are a weighted
sum of the coordinates of its vertexes, with non-negative weights that
sum to unity.  So by the fundamental equation (\ref{Pax}), if we treat
$P(F)$ as weights, the point given by the transition probabilities of
a system with transfer functions $F$ lies in the probability polytope
whose vertexes are defined by these transfer functions. This defines
the so called vertex- or ${\cal V}$-representation of the probability
polytope. 

\begin{figure}
\includegraphics[width=8cm]{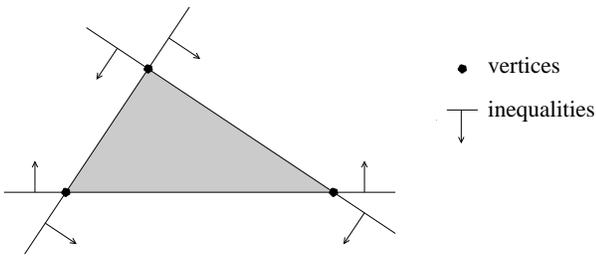}
\caption{\label{fig:tri}The two possible representations of a polytope by its vertexes (${\cal V}$-representation)
and by inequalities characterizing half-spaces
(${\cal H}$-representation).}
\end{figure}

There is an alternative representation of this probability polytope,
the so called half-space- or ${\cal H}$-representation, in terms of a
set of inequalities, each of which defines a half space.  A central
theorem of polytope theory \cite{Ziegler(1995)} states that any
polytope can always be described either by its ${\cal
V}$-representation or by its ${\cal H}$-representation.
Figure~\ref{fig:tri} illustrates this equivalence of the two possible
representations schematically in the simple case of a triangle.

\subsubsection{Local probability (Bell) polytopes\label{sub:local-polytopes}}

Let us consider the special case of local classical correlations in
more detail. These correlations are of central interest for the analysis of Bell
experiments that test locality.  Provided there are $N(x)$ and $N(y)$
possible values of inputs $x$ and $y$ and $N(a)$ and $N(b)$ possible
values for the outputs $a$ and $b$ of A and B, there are
$N(x)\times N(y)\times N(a)\times N(b)$ possible local transition
probabilities and ${N(a)}^{N(x)}\times {N(b)}^{N(y)}$ possible local
transfer functions.  Each of these transfer functions defines a vertex
of the corresponding local probability or Bell polytope ${\cal L}$,
thus yielding its ${\cal V}$-representation.  Any physical system
whose transition probabilities are located outside this Bell polytope
is nonlocal.

Whereas the ${\cal V}$-representation of a Bell polytope can be
obtained in a straightforward way from the transfer functions, the
determination of its corresponding ${\cal H}$-representation is a
considerably more difficult numerical problem for polytopes with large
numbers of vertexes \cite{Werner}. The half-spaces of the ${\cal
H}$-representation of a Bell polytope can be divided into several
classes. The least interesting class of inequalities expresses the
non-negativity conditions of all probabilities involved, i.e.
\begin{equation}
P(ab|x\y)\geq0\quad\forall\,(abx\y),\label{eq:positivity}\end{equation}
and the corresponding normalization conditions, i.e.
\begin{equation}
\sum_{ab}P(ab|x\y)=1\quad\forall\,(x\y)\,.\label{eq:normalization}
\end{equation}
A more interesting class of constraints are the
\emph{no-signaling equalities}, 
\begin{eqnarray}
\sum_{a}P(ab|x_{1}y)&=&\sum_{a}P(ab|x_{2}y)\quad\forall\,(x_{1}x_{2}yb),\nonumber\\
\sum_{b}P(ab|x\y_{1})&=&\sum_{b}P(ab|x\y_{2})\quad\forall\,(x\y_{1}y_{2}a),\label{eq:nosignalling}
\end{eqnarray}
which follow because for a local system no signals may be sent from
$A$ to $B$ or from $B$ to $A$.

The third and most interesting inequalities of the ${\cal
H}$-representation are the Bell inequalities themselves.  Because the
transformation from ${\cal V}$- to ${\cal H}$-representation is
difficult, they are known only in special cases, so the search for new
families of Bell inequalities is still an active research area
\cite{Werner,researchonBellineq1,researchonBellineq2,researchonBellineq3}.

Our subsequent discussion will focus on the ${\cal V}$-representation
of Bell polytopes, as they can be determined from the relevant
transfer functions in a straightforward way. Furthermore, by taking
into account constraints on the transition probabilities arising from
conservation of probability and from locality, the ${\cal V}$
-representation of Bell polytopes can be obtained efficiently in a
reduced basis. This fact was realized earlier already by Pitovski
\cite{Pitovski1,Pitovski2,Pitovski3}.  Let us start
from the observation that the full space of local transition
probabilities is $N(x)\times N(a)\times N(y)\times N(b)$-dimensional.
Due to conservation of probability these transition probabilities
fulfill the $N(x)\times N(y)$ relations
\begin{eqnarray}
\sum_{a,b}~P(ab|xy) &=&1.
\end{eqnarray}
Thus, for each choice of input $(x,y)$ by A and B one output,
say $(a_{N(a)}, b_{N(b)})$, can be eliminated by this linear
dependence.  Furthermore, if A's output is equal to $a_{N(a)}$ all
joint transition probabilities involving this output can be expressed
as
\begin{eqnarray}
P(a_{N(a)} b|xy) &=& P_B(b|xy) - \sum_{a\neq a_{N(a)}}P(ab|xy),
\label{margAlice}
\end{eqnarray}
where $P_B(b|xy) = \sum_{a}P(ab|xy)$ is B's marginal transition
probability.  Because of the no-signaling constraints
(\ref{eq:nosignalling}) this marginal transition probability cannot
depend on A's choice of input $x$, i.e. $P_B(b|xy) \equiv P_B(b|y)$.
An analogous argument applies to B, i.e.
\begin{eqnarray}
P(a b_{N(b)}|xy) &=& P_A(a|xy) - \sum_{b\neq b_{N(b)}}P(ab|xy)
\label{margBob}
\end{eqnarray}
with
$P_A(a|xy) \equiv P_A(a|x)$.
Thus, the marginal and joint transition probabilities which do not
contain the outputs $a_{N(a)}$ or $b_{N(b)}$ are linearly independent
and span the full space of local transition probabilities of a Bell polytope,
so they form a basis of dimension 
\begin{eqnarray}
D &=& N(x)\times (N(a)-1) + N(y)\times (N(b)-1) +\nonumber\\
&&N(x)\times N(y) \times (N(a)-1)\times (N(b) - 1)
\label{Dimension}
\end{eqnarray}
for the Bell polytope.  
As a result the Bell polytope is given by
all these linearly independent marginal and joint transition probabilities fulfilling the
conditions
\begin{eqnarray}
\sum_{a\neq a_{N(a)}} P_A(a|x) &\leq& 1,~
\sum_{b\neq b_{N(b)}} P_B(b|y) \leq 1, \nonumber\\
P(ab|xy) &=& P_A(a|x)P_B(b|y)
\label{Pitovskivertexes}
\end{eqnarray}
for $a\neq a_{N(a)},b\neq b_{N(b)}$ and all possible $N(x)\times N(y)$ inputs.
Furthermore, the vertexes of the Bell polytope are given 
by all those points in this probability
space whose coordinates assume the values $0$ and $1$ only and which are consistent
with relations (\ref{Pitovskivertexes}). 

\subsubsection{Nonlocal no-signaling probability polytopes\label{sub:non-signalling-polytope}}

In an experiment with detection efficiency close to the ideal, it
would be possible to demonstrate the correlations of weak nonlocality,
but so far there has always been at least one loophole
\cite{newtonlecture}.  Further, Bell's theorem itself
is incomplete, since it is based on the unproved assumption that such
detection is possible \cite{PercivalGarraway,Dirac}.

Thus, it is of interest to determine minimum detection efficiencies
which still allow for a violation of locality by quantum theory
provided the local measurements of A and B are separated by a
spacelike interval.  For the determination of these minimum detection
efficiencies a detailed knowledge of the set of correlations produced
by entangled quantum systems is required.  Unfortunately, a complete
characterization of all possible correlations of bipartite local
measurements of quantum systems does not yet exist \cite{Acin}.

However, for any given number of inputs and outputs nonlocal
no-signaling polytopes ${\cal P}$ can be constructed which include all
possible bipartite quantum correlations of quantum systems of
arbitrary dimensions.  Therefore, the boundary of the region 
representing all possible bipartite quantum correlations is sandwiched
between the boundaries of a nonlocal no-signaling polytope ${\cal P}$ and
its corresponding Bell polytope.
For any given set of inputs and outputs, this enables one to obtain
lower bounds on minimum detection inefficiencies which still allow the
observation of nonlocal features of quantum systems and which are
independent of the dimension of the quantum system and the associated
choice of quantum observables generating these correlations.

In the case of $N(x)\times N(y)$ inputs and $N(a)\times N(b)$ outputs,
the associated nonlocal no-signaling polytope is defined by all joint
transition probabilities $P(ab|xy)$ which are constrained only by the
no-signaling conditions (\ref{eq:nosignalling}). It should be stressed
that these no-signaling conditions are weaker than locality because
they do not necessarily imply that the underlying transfer functions
are local.

Thus, in general the no-signaling conditions are also compatible with
transfer functions of the form $F\neq (F^A,F^B)$.  As with the local
Bell polytopes of Sec.~\ref{sub:local-polytopes}, the nonlocal
no-signaling polytopes can be described conveniently in the reduced
basis formed by all marginal and joint transition probabilities which
do not contain outputs $a_{N(a)}$ or $b_{N(b)}$ and whose dimension is
given by relation (\ref{Dimension}).  In this reduced basis the
no-signaling constraints (\ref{eq:nosignalling}) are already taken
into account provided the marginal and joint transition probabilities
fulfill the consistency constraints
\begin{eqnarray}
\sum_{a\neq a_{N(a)}}P(ab|xy) \leq P_B(b|y),\nonumber\\
\sum_{b\neq b_{N(b)}}P(ab|xy) \leq P_A(a|x)
\end{eqnarray}
for all inputs $(x,y)$.  These inequalities follow from
Eqs. (\ref{margAlice}) and (\ref{margBob}) and the no-signaling constraints (\ref{eq:nosignalling}). 
So the requirement of no-signaling is weaker than locality.
Furthermore, these inequalities indicate that the nonlocal
no-signaling polytopes are defined in a natural way in the ${\cal
H}$-representation.  Thus, for large dimensions of the reduced basis
obtaining the corresponding ${\cal V}$-representation from this ${\cal
H}$-representation is a difficult numerical problem that limits the number of inputs and outputs 
considerably for
which this conversion can be achieved.

\section{Detection inefficiencies\label{Detineff}}

Based on the previously discussed local and no-signaling polytopes
${\cal L}$ and ${\cal P}$, in this section lower bounds on minimum
detection efficiencies are obtained below which violations of locality
cannot be observed in bipartite Bell experiments.  The dependence of
these lower bounds on the numbers of inputs and outputs and on
symmetry is explored.  It should be emphasized that for a given number
of inputs and outputs these lower bounds on minimum detection
efficiencies apply to correlations originating from arbitrary
bipartite quantum states and observables of arbitrary dimensional
Hilbert spaces.  A related problem, namely the determination of
maximum possible values of detection efficiencies which still
guarantee locality, has recently been investigated by Bigelow
\cite{Bigelow} with the help of linear programming techniques for some
special cases of correlations originating from two- and three-qubit
systems. Contrary to our approach this investigation does not involve
the nonlocal no-signaling polytope so that its resulting conclusions
apply only to correlations which originate from two- and three-qubit
quantum systems and from particular choices of quantum observables.

One of the simplest ways to describe detection inefficiencies of A and
B is by a parameter $\eta\in[0,1]$ describing the total efficiency of
the detection systems involved. Thus, $\eta$ is the probability that a
detector fires if it actually should.  In practice, detection
inefficiencies can have different physical origins.  They can
originate from an imperfect photodetector, for example, which does not
respond to each photon hitting its detection surface. Alternatively,
they may also arise from the fact that due to the particular geometry
of an experimental setup only a fraction of photons propagating within
a small solid angle is capable of hitting a photodetector at all. In
the following we assume that a combination of these effects gives rise
to the detection efficiencies $\eta_1$ and $\eta_2$ of A and B in a
bipartite Bell experiment. Furthermore, these detection efficiencies
are assumed to be independent of the polarization of the photons
hitting the photodetectors.

For ideal detection, in a dichotomic Bell experiment A always receives
a photon, so she needs only one detector to distinguish the
polarizations.  But for real detectors, a single
polarization-sensitive detector makes no distinction between the
absence of a photon and a photon with the wrong polarization, whereas
two polarization-sensitive detectors can distinguish between these two
cases.  Similarly for B. Thus, in imperfect situations two detectors
give output events that are not possible with only one, increasing the
dimension of the relevant polytopes. Indeed, we will demonstrate that
the two cases can give rise to different lower bounds on detection
efficiencies.

First of all let us describe detection inefficiencies where A's (B's)
detector cannot distinguish between the no-detection event and the
event $a_{N(a)}$ ($b_{N(b)})$). Thus, the ideal joint transition
probabilities, $P_{1,1}(ab|xy)$, are related to the corresponding
imperfect joint transition probabilities, $P_{\eta_1,\eta_2}(ab|xy)$,
by
\begin{eqnarray}
P_{\eta_1,\eta_2}(ab|xy) & = & \eta_1 \eta_2 P_{1,1}(ab|xy),\nonumber\\
P_{\eta_1,\eta_2}(ab_{N(b)}|xy) & = & \eta_1  P_{1,1}(ab_{N(b)}|xy) +\nonumber\\
&& \eta_1 (1 - \eta_2) \sum_{b\neq b_{N(b)}}P_{1,1}(ab|xy),\nonumber\\  
P_{\eta_1,\eta_2}(a_{N(a)}b|xy) & = & \eta_2  P_{1,1}(a_{N(a)}b|xy) +\nonumber\\
&&\eta_2 (1 - \eta_1) \sum_{a\neq a_{N(a)}}P_{1,1}(ab|xy),\nonumber\\  
P_{\eta_1,\eta_2}(a_{N(a)}b_{N(b)}|xy) & = & \eta_1\eta_2 P_{1,1}(a_{N(a)}b_{N(b)}|xy) +\nonumber\\
&&\eta_1 (1 - \eta_2) \sum_{b}P_{1,1}(a_{N(a)}b|xy) +\nonumber\\
&&\eta_2 (1 - \eta_1) \sum_{a}P_{1,1}(a b_{N(b)}|xy) +\nonumber\\
&&(1-\eta_1)(1-\eta_2)
\label{merge1}
\label{relations}
\end{eqnarray}
for all outputs $(a\neq a_{N(a)}, b\neq b_{N(b)})$ and inputs $(x,y)$
of A and B.  For dichotomic Bell experiments with photons this
describes situations in which A and B each use one
polarization-sensitive photodetector only which cannot distinguish
between a photon with the wrong polarization and a no-detection event.
In the reduced basis of marginal and joint transitions probabilities
discussed in Secs.~\ref{sub:local-polytopes} and
\ref{sub:non-signalling-polytope}, in which the outputs $a_{N(a)}$ and
$b_{N(b)}$ are eliminated, these relations reduce to the simple form
\begin{eqnarray}
P_{A\eta_1\eta_2}(a|x) &=& \eta_1 P_{A1}(a|x),~P_{B\eta_1\eta_2}(b|y) = \eta_2 P_{B1}(b|y),\nonumber\\
P_{\eta_1\eta_2}(ab|xy) &=& \eta_1\eta_2 P_{1,1}(ab|xy)
\label{merge2}
\end{eqnarray}
for all
outputs $(a\neq a_{N(a)}, b\neq b_{N(b)})$ and inputs $(x,y)$ of A and B.

If, in contrast, the no-detection event $\emptyset$ is treated as an
additional output, the dimension of the relevant transition
probability polytope is increased. In this case the ideal and
imperfect transition probabilities $P_{1,1}(ab|xy)$ and $P_{\eta_1
\eta_2}(ab|xy)$ are related by
\begin{eqnarray}
P_{\eta_1,\eta_2}(ab|xy) & = & \eta_1 \eta_2 P_{1,1}(ab|xy),\nonumber\\
P_{\eta_1,\eta_2}(a\emptyset|xy) & = & \eta_1 (1 - \eta_2) \sum_{b}P_{1,1}(ab|xy),\nonumber\\  
P_{\eta_1,\eta_2}(\emptyset b|xy) & = &  (1 - \eta_1) \eta_2 \sum_{a}P_{1,1}(ab|xy),\nonumber\\  
P_{\eta_1,\eta_2}(\emptyset \emptyset|xy) & = & (1 -\eta_2) (1 - \eta_1) 
\label{emptyset}
\end{eqnarray}
for all outputs $(a,b)$ and inputs $(x,y)$ of A and B.  For dichotomic
Bell experiments with photons this describes situations in which A and
B each use two photodetectors which are sensitive to two orthogonal
polarizations.  By eliminating from Eqs. (\ref{emptyset}) all joint
transition probabilities involving the outputs $a_{N(a)}$ or
$b_{N(b)}$ with the help of the marginal transition probabilities one
obtains the corresponding relations between the ideal and imperfect
transition probabilities of the reduced basis.

For a given number of inputs and outputs, lower bounds on detection
efficiencies below which a violation of locality is no longer possible
can be obtained from Eqs. (\ref{merge1}), (\ref{merge2}), and
(\ref{emptyset}) by identifying the ideal transition probabilities
$P_{1,1}(ab|xy)$ with the possible correlations of the nonlocal
no-signaling polytope ${\cal P}$ and by determining the critical
detection efficiencies $(\eta_1,\eta_2)$ at which the corresponding
imperfect transition probabilities $P_{\eta_1,\eta_2}(ab|xy)$ merge
into the Bell polytope ${\cal L}$.

These critical detection efficiencies $\eta_1$ and $\eta_2$ determine
lower bounds on the detection efficiencies below which a violation of
locality is no longer possible by the corresponding correlations
produced by any quantum system.  In general, it is unclear whether the
lower bounds obtained on the basis of the no-signaling polytope ${\cal
P}$ can be reached by any quantum system with appropriate choices of
the dimension of the Hilbert space and of the quantum observables. But
it is shown later that in the special case of two inputs and two
outputs of both A and B these lower bounds actually can be reached.

\begin{table}
\begin{tabular}{|c||c|c|c|c|}
\hline $\left(A,B\right)$ & $\eta_{1}=\eta_{2}$ & $\eta_{1}=\eta_{2}$
& $\eta_{1}=1$ & $\eta_{1}=1$\tabularnewline   & add. 
& no add.  & add.  & no add. \tabularnewline
\# inputs &  outcome
&  outcome &  outcome &  outcome\tabularnewline
\hline \hline $\left(2,2\right)$ & 0.6667 & 0.6667 & 0.5000 &
0.5000\tabularnewline \hline $\left(3,2\right)$ & 0.6667 & 0.6667 &
0.5000 & 0.5000\tabularnewline \hline $\left(3,3\right)$ & 0.5714 &
0.6000 & 0.3333 & 0.3333\tabularnewline \hline $\left(4,3\right)$ &
0.5000 & 0.5714 & 0.2500 & 0.2500\tabularnewline \hline
\end{tabular}
\caption{Critical detection efficiencies, $\eta_1$ and $\eta_2$, of Alice (A)  and Bob (B)
for dichotomic bipartite symmetric ($\eta_1=\eta_2$) and extreme asymmetric ($\eta_1 =1$)
Bell experiments with
various numbers of inputs and for cases in which the no-detection event is treated separately
(add. outcome) and in which it is combined with the event $(a_2,b_2)$ (no add. outcome).\label{Table1}}
\end{table}

Table \ref{Table1} summarizes numerically-determined lower bounds on
detection efficiencies $(\eta_1$ and $\eta_2)$ of A and B which
characterize the merging of the imperfect transition probabilities
$P_{\eta_1,\eta_2}$ into the relevant local polytope ${\cal L}$.  If A
and B randomly choose one of two possible physical variables in their
respective laboratories, i.e. case (2,2), and if they have identical
detectors, i.e. $\eta\equiv\eta_1 = \eta_2$ (symmetric case), the
resulting critical value of $\eta$ turns out to be independent of
whether or not the no-detection event is treated as an additional
output.  This optimal lower bound obtained on the basis of the
no-signaling polytope ${\cal P}$ turns out to be identical with the
minimal detection efficiency obtained previously by Eberhard
\cite{Eberhard}.  Eberhard's result demonstrated that pure two-qubit
quantum states exist which are able to violate locality down to
minimum detection efficiencies of magnitude $\eta= 0.6667$ in
symmetric cases. 
Surprisingly these most robust two-qubit quantum
states are almost separable.  Also the optimal lower bounds for the
corresponding extreme asymmetric cases of Table \ref{Table1},
i.e. $\eta_1=1 \neq \eta_2$ of magnitude $\eta_2 = 0.5000$, in which
Alice's detection efficiency is assumed to be perfect, are independent
of whether or not the no-detection event is treated as an additional
output. These lower bounds also agree with the minimum possible
detection efficiencies of Eberhard \cite{Eberhard} which just allow
for a violation of locality by two-qubit quantum systems.

It should be mentioned that apart from reproducing Eberhard's previous
minimum detection efficiencies our results of Table \ref{Table1} for
the $(2,2)$-case also demonstrate that there is no way to violate
locality with detection efficiencies below $\eta = 0.667$ in the
symmetric case and below $\eta_2 = 0.500$ in the extreme asymmetric
case. This conclusion holds for arbitrary choices of two two-valued
quantum observables of A and B and for arbitrary bipartite quantum
states in arbitrary dimensional Hilbert spaces, which produce the
statistical correlations.

Table \ref{Table1} also includes results on optimal lower bounds for
cases in which more than two physical variables are selected on
Alice's or Bob's sides. It is apparent that in general these lower
bounds depend on whether or not the no-detection event is treated as
an additional output.  Furthermore, the lower bounds of cases in which
the no-detection event is treated as an additional output are always
lower than or equal to cases in which the no-detection event is
combined with an already existing output. 
However, for cases with
more than two inputs of Alice or Bob it is not known yet whether
quantum systems exist which are capable of violating locality all the
way down to these lower bounds.  However, the number of outputs on
Alice's and Bob's sides, $N(a)$ and $N(b)$, puts a lower bound on the
dimension $D$ of the Hilbert space of these quantum systems,
i.e. $D\geq N(a)\times N(b)$.

\begin{figure}
\includegraphics[width=8cm]{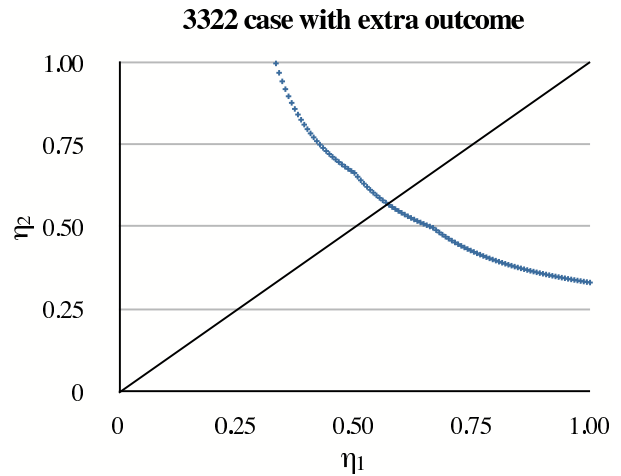}
\caption{Lower bounds on detection efficiencies $(\eta_1,\eta_2)$ for three inputs on Alice's and Bob's sides
with the no-signal event treated as an extra output.\label{fig:eta1eta2a}}
\end{figure}

\begin{figure}
\includegraphics[width=8cm]{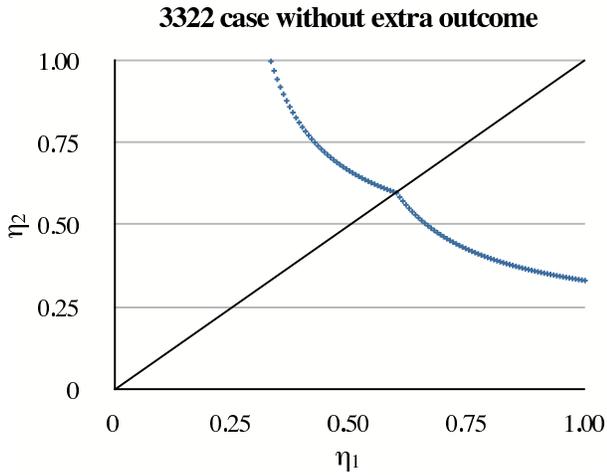}
\caption{Lower bounds on detection efficiencies $(\eta_1,\eta_2)$ for three inputs on Alice's and Bob's sides
with the no-signal event combined with the output $(a_2,b_2)$.\label{fig:eta1eta2b}}
\end{figure}

In Figs.~\ref{fig:eta1eta2a} and \ref{fig:eta1eta2b} lower bounds on
minimum detection efficiencies $(\eta_1,\eta_2)$ are depicted for
arbitrary cases between the symmetric $(\eta_1=\eta_2$) and the
extreme asymmetric ($1=\eta_1\neq \eta_2$) situation for the special
case of three inputs and two outputs of both A and B.  Irrespective
whether or not the no-signal event is treated as a separate outcome
one observes a cusp-like dependence in these figures.  This non-smooth
dependence corresponds to a case in which, at a particular value of
$(\eta_1,\eta_2)$, a vertex of the properly transformed nonlocal
no-signaling polytope (compare with Eqs.(\ref{merge1})) just coincides
with a vertex of the local (Bell) polytope ${\cal L}$.

We have also explored lower bounds on detection efficiencies for two
inputs and three outputs on both Alice and Bob's sides. In the
symmetrical case ($\eta_1=\eta_2\equiv \eta$) the lower bound was
given by $\eta=0.6667$,  whether or not the no-signal
event was combined with an output.  Similarly for the extreme
asymmetric case ($\eta_1=1$) we obtained the lower bound $\eta_2 =
0.5000$.

It is difficult to determine lower bounds on detection efficiencies
numerically with the help of the no-signaling polytope ${\cal P}$ for
larger numbers of inputs or outputs. This is due to the fact that
no-signaling polytopes are defined in a natural way in the ${\cal
H}$-representation (compare with the discussion of
Sec.~\ref{sub:non-signalling-polytope}). Thus in order to determine
lower bounds on detection efficiencies one has to convert the nonlocal
no-signaling polytope from its ${\cal H}$-representation into its
${\cal V}$-representation, which becomes very difficult numerically
for such cases.


\section{Distance measures quantifying the violation of locality\label{Distmeas}}

\begin{figure}
\includegraphics[width=8cm]{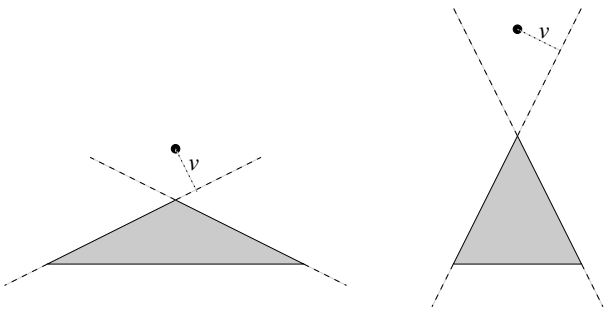}
\caption{\label{fig:violations}Schematic representations of identical violations $v$ of
relevant Bell inequalities. Note the different distances from the polytopes.}
\end{figure}

The simplest way of testing correlations for locality is to determine
whether or not the relevant point ${\bf X}$ of transition-probability space
is inside the local polytope ${\cal L}$. However, in general
probabilities can be estimated experimentally only approximately with
an uncertainty depending on the size of the statistical sample
involved.  Thus, a practically useful method for determining
violations of locality should also quantify how far outside
${\cal L}$ the given point ${\bf X}$ is located.  Therefore, it is desirable
to develop methods which permit one to find the distance of a given
point ${\bf X}$ in transition-probability space from a local polytope ${\cal
L}$, so that one can decide whether observed transition probabilities
with their experimental uncertainties still violate locality.

A natural choice for such a distance measure is the Euclidean $2$-norm
distance from the nearest facet of the polytope ${\cal L}$ which
corresponds to a Bell inequality.  But also the $1$- or the
$\infty$-norms are possible choices.  However, as illustrated in
Fig. \ref{fig:violations}, it definitely makes more sense to consider
the ``distance of ${\bf X}$ from the polytope ${\cal L}$ as a whole''.  The
only reasonable definition for the ``distance from a polytope'' is the
minimum distance of any point of ${\cal L}$ to ${\bf X}$, which means we
have to deal with an optimization problem.

Let us assume ${\cal L}$ is given in $D$-dimensional space and that it
has $r$ vertexes, which we denote by ${\bf v}_i\in\mathbb{R}^{D}$.
Thus, any point ${\bf Y}\in {\cal L}$ can be written as a convex
combination of these vertexes, i.e.  ${\bf Y}=\sum_{i=1}^{r}w_{i}{\bf
v}_{i}$ with weights $w_{i}\geq0$ and with $\sum_{i=1}^{r}w_{i}=1$.
So, it is natural to use the $w_{i}$ (or the vector ${\bf w}^T =
(w_1;w_2;\ldots;w_r)$) as the coordinates for the optimization problem
rather than the coordinates of ${\bf Y}$ in the actual space
$\mathbb{R}^{D}$ in which the polytope lives.  This is motivated by
the fact that the constraints of the polytope are given in terms of
the weights $w_{i}$ rather than in terms of the coordinates of the
actual space.  However, we still want to optimize the distance in the
actual space.  In order to achieve this for the $1$-norm, it
is convenient to introduce the matrix ${\bf C}=({\bf v}_{1};{\bf
v}_{2};\ldots;{\bf v}_{r})\in\mathbb{R}^{D\times r}$ with ${\bf Y} =
{\bf C}\cdot {\bf w}$.  Let us also use the abbreviations
$\boldsymbol{0}_{D}$, $\boldsymbol{1}_{D}$ and $\boldsymbol{0}_{r}$,
$\boldsymbol{1}_{r}$ for column vectors of all zeros or ones in
$\mathbb{R}^{D}$ and $\mathbb{R}^{r}$, respectively. Analogously we
use the notation ${\bf 1}_{D\times D}$ for a diagonal $D\times D$ unit
matrix, and similarly ${\bf 1}_{r\times r}$.  The problem of finding the minimum distance between a point
${\bf X}\in {\mathbb R}^D$ and the local polytope ${\cal L}$ can
now be formulated as the following linear programming problem:
\begin{quote}
Maximize $-(\boldsymbol{1}_{D}^T;\boldsymbol{0}_{r}^T)\cdot{\bf Z}$\\
Subject to ${\bf A}\cdot{\bf Z} \leq {\bf b}$
\end{quote}
with the $(2D+3)\times (D+r)$ matrix
\begin{eqnarray}
{\bf A} &=&\left(
\begin{array}{ccr}
-{\bf 1}_{D\times D}&,&{\bf C}\\
-{\bf 1}_{D\times D}&,&{\bf C}\\
\boldsymbol{0}_{D}^T&,&-\boldsymbol{1}_{r\times r}\\
\boldsymbol{0}_{D}^T&,&\boldsymbol{1}_{r}^T\\
\boldsymbol{0}_{D}^T&,&-\boldsymbol{1}_{r}^T
\end{array}
\right),
\end{eqnarray}
the $(D+r)$-dimensional vector ${\bf Z}^T = (\bar{\bf Z}^T;{\bf w}^T)$
and the $(2D+r+2)$-dimensional vector
${\bf b}^T = ({\bf X}^T;-{\bf X}^T;0;1,;1)$. 
As a result
the $1$-norm is given by $(\boldsymbol{1}_{D}^T;\boldsymbol{0}_{r}^T)\cdot{\bf Z} \equiv \boldsymbol{1}_{D}^{T}\cdot\bar{\bf Z}$.

\begin{figure}
\includegraphics[width=8cm]{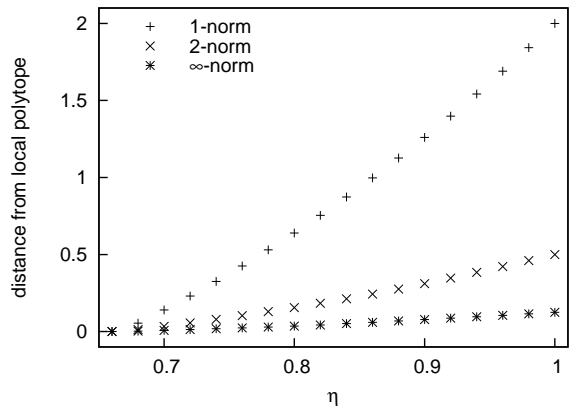}
\caption{\label{fig:distances}Distance of a vertex of ${\cal P}$  from ${\cal L}$ as a function of $\eta$
for two inputs and outputs of both A and B.}
\end{figure}

A similar linear programming problem can be formulated in order to
find the $\infty$-norm.  Although an analogous quadratic programming
problem can be formulated for the ordinary $2$-norm distance, it is
worth mentioning that the numerical solution of this quadratic problem
is much more difficult and time-consuming than the corresponding
linear programming problem.  The $2$-norm can however be bounded from
above and below by the $1$-norm and $\infty$-norm, respectively. As an
example, consider Fig.~\ref{fig:distances} which shows how the
distance between a properly transformed vertex of the nonlocal
no-signaling polytope ${\cal P}$ (compare with Eqs.~(\ref{merge1}))
and the local polytope ${\cal L}$ varies smoothly when we vary the
detection efficiency.  Of course, at $\eta=0.6667$ all of the distance
measures vanish.

\section{Summary and Conclusions}

For given numbers of inputs and outputs we have investigated minimum
detection efficiencies below which locality cannot be violated by
correlations produced by any quantum system in bipartite Bell
experiments. For this purpose lower bounds on these minimum detection
efficiencies have been obtained numerically with the help of linear
programming techniques.  Our determination of these lower bounds is
based on the observation that for any given number of inputs and
outputs any possible bipartite correlation produced by a quantum
system in an arbitrary dimensional Hilbert space is sandwiched between
the boundaries of the nonlocal no-signaling polytope and the Bell
polytope.
Thus, for imperfect detection the detection efficiencies at which
statistical correlations of the properly transformed nonlocal
no-signaling polytope merge into the Bell polytope yield lower bounds
on these minimum detection efficiencies.

Both the local (Bell) and the nonlocal no-signaling polytope can can
be dealt with conveniently by linear programming.  In particular, the
vertex representation of any Bell polytope can be determined in a
straightforward way. The construction of the nonlocal no-signaling
polytope is more complicated as it is naturally defined in the ${\cal
H}$-representation.

Our numerically calculated lower bounds on detection efficiencies
demonstrate that in general, with the exception of two inputs and
outputs of A and B, these bounds are not identical for Bell
experiments with symmetric and asymmetric detection efficiencies.
Furthermore, in the case of two inputs and outputs our lower bounds
agree with the minimum detection efficiencies obtained previously by
Eberhard \cite{Eberhard} for two-qubit quantum correlations. Thus, in
this case our results demonstrate that these minimum detection
efficiencies cannot be lowered even if one considered quantum
correlations originating from quantum systems of arbitrary dimensions.

Our investigation constitutes a first step towards a systematic study
of bipartite correlations produced by quantum systems.  In general, it
is still unclear to what extent our numerically determined lower
bounds can be reached by correlations of appropriately chosen quantum
systems.  Further research is required to clarify this point. In
addition, for numerical purposes it would be desirable to find an
effective way for determining directly the ${\cal V}$-representations
of nonlocal no-signaling polytopes without involvement of their ${\cal
H}$-representations. This would also allow the efficient treatment of
cases involving many inputs and outputs.  Furthermore, our approach
can also be adapted to the investigation of multipartite correlations.

\section*{Acknowledgments}
Financial support by the DAAD is gratefully acknowledged.

\end{document}